\documentclass[apj]{emulateapj}

\def\ni{\noindent}

\begin{document}

\shorttitle{Galactic Cepheids Distance using the Wesenheit Function}
\shortauthors{Ngeow}

\title{On the Application of Wesenheit Function in Deriving Distance to Galactic Cepheids}

\author{Chow-Choong Ngeow}
\affil{Graduate Institute of Astronomy, National Central University, Jhongli City, 32001, Taiwan}

\begin{abstract}

In this work, we explore the possibility of using the Wesenheit function to derive individual distances to Galactic Cepheids, as the dispersion of the reddening free Wesenheit function is smaller than the optical period-luminosity (P-L) relation. When compared to the distances from various methods, the averaged differences between our results and published distances range from $-0.061$ to $0.009$, suggesting that the Wesenheit function can be used to derive individual Cepheid distances. We have also constructed Galactic P-L relations and selected Wesenheit functions based on the derived distances. A by-product from this work is the derivation of Large Magellanic Cloud distance modulus when calibrating the Wesenheit function. It is found to be $18.531\pm0.043$ mag.

\end{abstract}


\section{Introduction}

Period-luminosity (P-L, also known as the Leavitt Law) relations based on classical Cepheids in our Galaxy, referred as the Galactic Cepheids, are important in distance scale work as well as in constraining the stellar pulsation and evolution calculations. Determining the P-L relations for Galactic Cepheids requires the measurement of distance to individual Cepheids. In contrast to extra-galactic Cepheids where the distance to the host galaxy can be obtained via independent means, there are only a limited number of methods to determine distances to Galactic Cepheids \citep[for example, see][and reference therein]{sto83,fea87,wal88,wil91,fea99,dib02,fea03,fou03,tam03,nge04,fou07,tur10}. These methods include: (1) direct parallax measurements from {\it Hipparcos} and {\it Hubble Space Telescope (HST)}; (2) main-sequence fitting to the open clusters or associations that host Cepheids; (3) Baade-Wesselink (BW) expanding photosphere techniques that combining the integration of radial velocity and angular diameter displacements measured from surface brightness relations, interferometric measurements or semi-theoretical approach; and (4) statistical parallax method that utilizes motions of Cepheids along the Galactic plane. Once the distances to a number of Galactic Cepheids have been measured using these methods, or a combination of them, Galactic P-L relations and the period-luminosity-color (P-L-C) relation can be calibrated. 

The calibrated P-L and P-L-C relations can then be used to derive distances to a larger number of Galactic Cepheids. This in turn can be used, for example, to investigate the structure and kinematics of our Galaxy \citep[see, for example,][]{sti56,kra63,fer68,tak70,efr81,cal87,opo88,pon94,zhu99,maj09} and mapping out the Galactic metallicity gradient or distribution \citep[see, for example,][]{gir83,and02,and04,kov05,luc06,yon06,lem07,lem08,ped09,luc11a,luc11}. However, values of extinction need to be assumed or adopted for the individual Cepheids before deriving their distances using the P-L and/or P-L-C relations. 

In this work, we examine the possibility of using the Wesenheit function \citep{bro80,mad82,mof86,mad91,kov97,cap00,kov01,leo03,nge05,fio07,bon08,bon10} to derive distances to individual Galactic Cepheids. This is because the intrinsic dispersion of P-L relations is on the order of $\sim0.2$ mag in the optical, hence the distance measured from using the P-L relations will suffer a systematic error of the same order. In contrast, the dispersion of the Wesenheit function is greatly reduced \citep{fio07,mad09}, as has been shown from the Large Magellanic Cloud (LMC) Cepheids: it is $\sim2$ to $\sim3$ times smaller as compared to the optical P-L relations \citep{tan99,uda99,fou07,sos08,nge09}. This can reduce the systematic error in derived distances. Furthermore, in order to correct for extinction, application of both the P-L and P-L-C relations require the estimation or determination of $E(B-V)$ values for each Cepheids. On the other hand, the Wesenheit function is reddening-free by construction \citep{fre88,mad91}. Hence the total systematic error of the derived distance does not include the extinction error when using the Wesenheit function. 

Using the Wesenheit function to derive distance to Galactic Cepheids has been suggested by \citet{opo83}. The calibration of the Wesenheit function given in \citet{opo83}, however, was based on 19 Cepheids located in open clusters or associations by setting the distance modulus of Hyades to be $3.31$mag, with a rather large dispersion of $0.158$ mag. Since then, large sets of photometric data from modern CCD measurements become available for the Galactic Cepheids. In addition, better independent distance measurements to a larger number of Galactic Cepheids, using varies techniques as mentioned previously, are available in the literature \citep[for example, high quality parallax measurements are available for $10$ Cepheids based on {\it HST} observations, see][]{ben07}. Therefore, the goal of this work is to examine the use of the Wesenheit function in deriving distances to Galactic Cepheids by taking advantage of these latest developments \citep[see][for a similar approach]{fio07}.

\section{Distances to Galactic Cepheids Using the Wesenheit Function}

The Wesenheit function can be defined in various bandpass and filter combinations \citep[for example, see][and reference therein]{nge05}. In this work, we only adopt the Wesenheit function in the form of $W=I-1.55(V-I)$, because it has been demonstrated that the $VI$ band based Wesenheit function is insensitive to metallicity \citep[see, for example,][]{maj11}, although an opposite result has been found in \citet{sto11b}. Nevertheless, as discussed in Section 4, we assume the adopted Wesenheit function is insensitive to metallicity. The Wesenheit function used in this work is adopted from \citet{nge09}: $W=-3.313\log(P)+15.892$, with a dispersion of $0.069$ mag. This Wesenheit function is derived from $\sim1500$ LMC fundamental mode Cepheids. Therefore, the intercept needs to be calibrated. This is equivalent to derive the distance modulus to LMC. Assuming that our Wesenheit function is applicable to Galactic Cepheids, then the $10$ Galactic Cepheids with parallax measurements from {\it HST} \citep{ben07} can used to calibrate the Wesenheit function. Based on these calibrators, the distance modulus of LMC was found to be $18.531\pm0.043$ mag. Therefore, distance moduli to Galactic Cepheids can be found using the following equation:

\begin{eqnarray}
\mu_W = I - 1.55(V-I) + 3.313\log(P)+2.639,
\end{eqnarray}

\ni where one only needs to know the pulsating period, $P$, and the $VI$ band intensity mean magnitudes (the $I$ band is in Cousin system) for a given Cepheid. Error on $\mu_W$ is taken to be $\sigma_W=\sqrt{(0.069)^2+(0.043)^2}=0.081$ mag. 

\begin{deluxetable*}{llcrrrrrrrccr}
\tabletypesize{\scriptsize}
\tablecaption{Distance to Galactic Cepheids Using the Wesenheit Function\tablenotemark{a}. \label{tab1}}
\tablewidth{0pt}
\tablehead{
\colhead{Name} &
\colhead{Type} & 
\colhead{$\log(P)$} &
\colhead{$B$} &
\colhead{$V$} &
\colhead{$R$} &
\colhead{$I$} &
\colhead{$J$} &
\colhead{$H$} &
\colhead{$K$} &
\colhead{$E(B-V)$} &
\colhead{$[\mathrm{Fe/H}]$} &
\colhead{$\mu_W$} 
}
\startdata
U AQL	&	DCEP &	0.846 &	7.477 & 6.430 &	5.829 & 5.278 &	999.99 & 999.99 & 999.99 & 0.381 & 0.17 & 8.934 \\
SZ AQL	&	DCEP &	1.234 &	10.041& 8.630 &	7.824 & 7.063 &	5.892  & 5.369  & 5.149  & 0.559 & 0.24 & 11.361 \\
TT AQL	&	DCEP &	1.138 &	8.424 & 7.131 &	6.410 & 5.718 &	4.690  & 4.208  & 4.014  & 0.493 & 0.22 & 9.937  \\
$\cdots$ & $\cdots$ & $\cdots$ & $\cdots$ & $\cdots$ & $\cdots$ & $\cdots$ & $\cdots$ & $\cdots$ & $\cdots$ & $\cdots$ & $\cdots$ & $\cdots$ 
\enddata
\tablenotetext{a}{$999.99$ indicates no data for a given entry. \\
(This table is available in its entirety in a machine-readable form in the online journal. A portion is shown here for guidance regarding its form and content.)
}
\end{deluxetable*}

A sample of Galactic Cepheids that have $VI$ band intensity mean magnitudes was compiled from \citet{ber00}, with updated intensity mean magnitudes adopted from \citet{van07}. Additional Cepheids or the missing $VI$ band intensity mean magnitudes were added from the following sources: \citet{gro99}, \citet{lan99}, \citet{san99}, \citet{tam03}, \citet{fou07} and \citet{van07}. In addition, $B$- and $R$-band (in Cousin system) intensity mean magnitudes were available for $387$ and $334$ Cepheids, respectively, from the cited reference. Finally, $JHK$ band intensity mean magnitudes in SAAO (South African Astronomical Observatory) system are available for $229$ Cepheids from \citet{van07}. They were converted to 2MASS \citep[Two-Micron All Sky Survey,][]{skr06} systems using the color transformation equations given on the 2MASS Web site\footnote{{\tt http://www.ipac.caltech.edu/2mass/releases/allsky/\\ doc/sec6\_4b.html}, which are updated transformation equations from \citet{car01}.}. According to GCVS \citep[General Catalog of Variable Stars,][]{sam09}, our sample includes $322$ Cepheids that are classified as DCEP\footnote{Both V1154 Cyg \citep{sza11} and FF Aql \citep{mar10} are updated to DCEP in this work.}, $34$ of them are classified as CEP and $37$ are classified as DCEPS. The $BVRIJHK$ band intensity mean magnitudes and the distance moduli calculated from equation (1) for these Cepheids are summarized in Table \ref{tab1}.

Extinction and metallicity for these Cepheids are also compiled in Table \ref{tab1} when available. For extinction corrections, $E(B-V)$ values taken from \citet{fer95}\footnote{Available at {\tt http://www.astro.utoronto.ca/DDO/research/cepheids/}} were scaled with a scale factor suggested by \citet[][that is, $E(B-V)=0.951\times E(B-V)_{\mathrm{Fernie}}$]{tam03}. $E(B-V)$ for 10 Cepheids, which do not have the entries from \citet{fer95}, were taken from either \citet{fou07} or \citet{van07}. The final adopted $E(B-V)$ values were listed in Table \ref{tab1}. For metallicity, $[\mathrm{Fe/H}]$ values are available from \cite[][$254$ Cepheids]{luc11}, \citet[][$76$ Cepheids]{luc11a}, \citet[][VW Cen \& LS Pup]{rom08}, \citet[][X Sgr]{and03}, \citet[][HQ Car]{yon06} and \citet[][QZ Nor]{fry97}. Metallicities from other publications are transformed to \citet{luc11} system by calculating the mean difference of the metallicity for common Cepheids, and the results are summarized in Table \ref{tab1}. The transformed metallicity and those from \citet{luc11} are listed in Table \ref{tab1} (when available). Metallicity for Polaris is available in \citet{and94}, thought it is not included in Table \ref{tab1}.  

\subsection{Comparison to Published Results}

\begin{figure*}
\plottwo{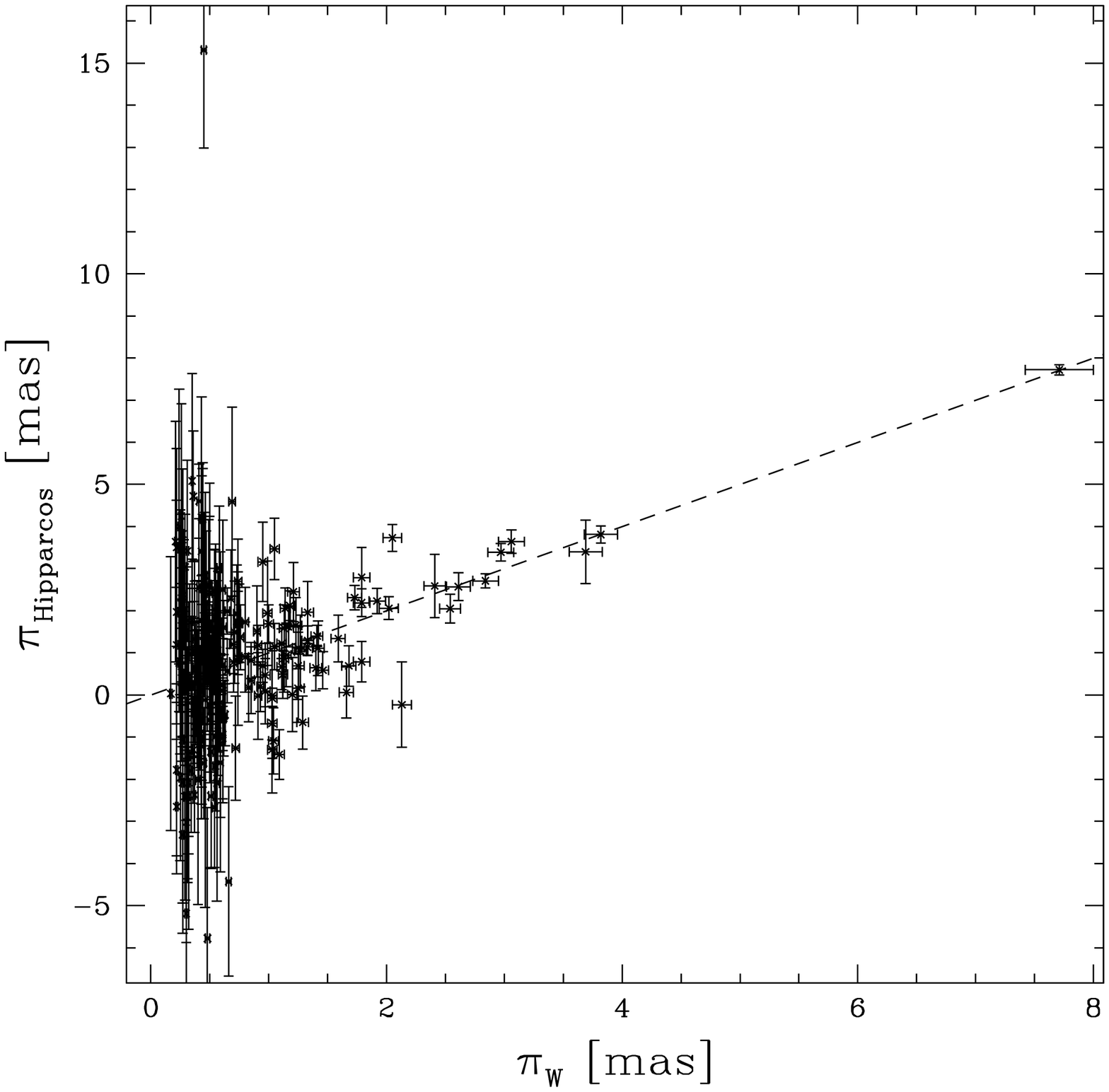}{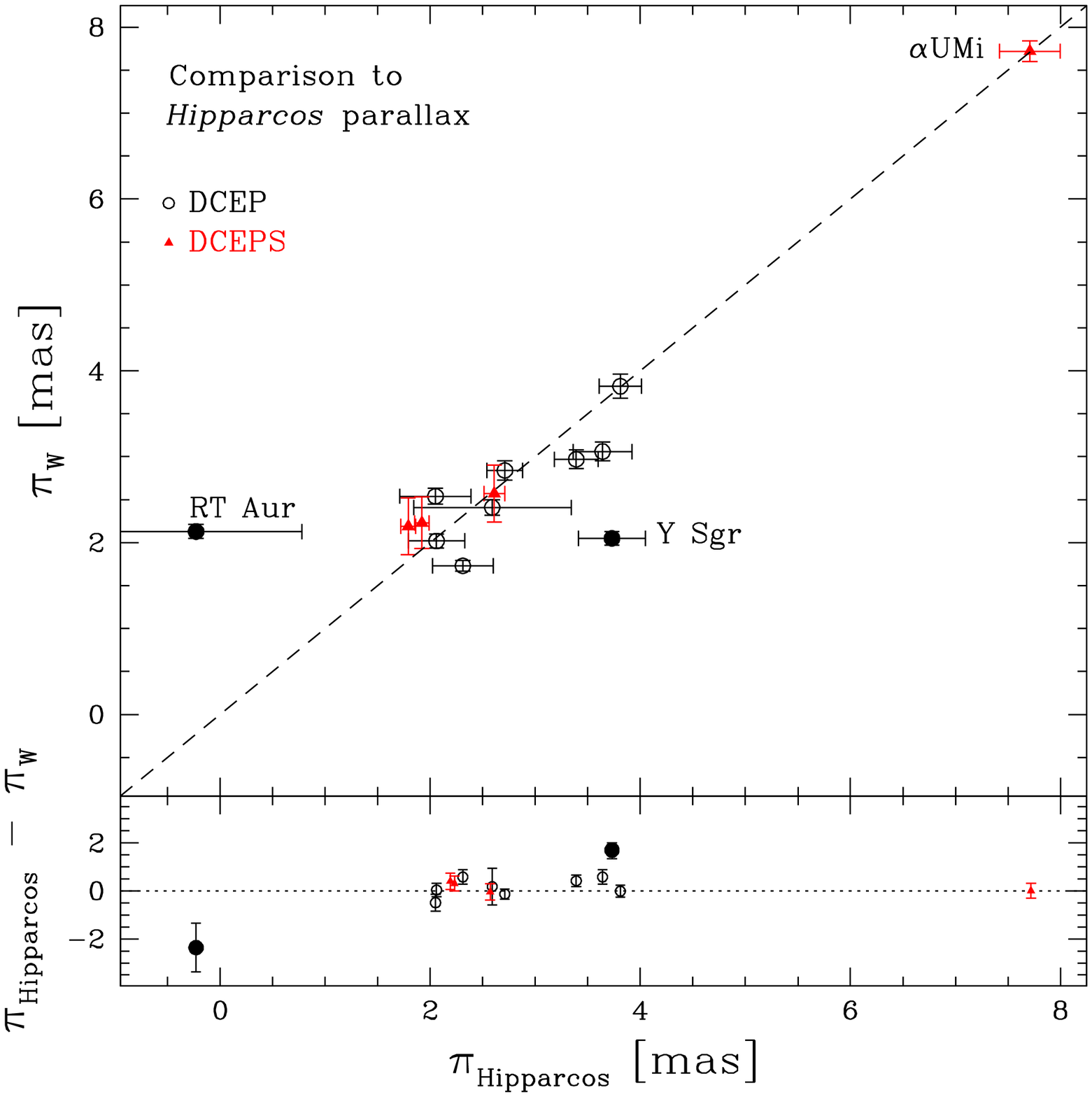}
\caption{Comparison of the {\it Hipparcos} parallaxes with parallaxes converted from distance moduli calculated using equation (1), denoted as $\pi_W$, for $223$ common Cepheids in the two samples (left panel) and for the Cepheids listed in Table 2 of \citet{van07}. DCEP and DCEPS Cepheids are represented by open circles and filled triangles, respectively. The dashed and dotted lines in the plots are for the case of $y=x$ and $y=0$, respectively, and not the fit to the data. Two discrepant Cepheids are marked with filled circles. [See on-line edition for color version of this Figure.] \label{fig_hipp}}
\end{figure*} 

To validate the use of Wesenheit function in deriving distance to individual Galactic Cepheids, distance moduli calculated from equation (1) can be compared to Galactic Cepheids that possess independent distance measurements available recently in literature. 

\begin{deluxetable}{lccc}
\tabletypesize{\scriptsize}
\tablecaption{Mean Dfference of Metallicity between Other Publication and \citet{luc11}. \label{tab2}}
\tablewidth{0pt}
\tablehead{
\colhead{Data set} &
\colhead{$N_{\mathrm{common}}$} & 
\colhead{$<\Delta>$\tablenotemark{a}} &
\colhead{$\sigma$\tablenotemark{b}} 
}
\startdata
\citet{luc11a} & 180 & 0.07 & 0.08 \\
\citet{and03}  & 48  & 0.08 & 0.08 \\
\citet{rom08}  & 25  & 0.11 & 0.11 \\
\citet{yon06}\tablenotemark{c}  & 18  & 0.28 & 0.13 \\
\citet{fry97}\tablenotemark{d}  & 10  & 0.19 & 0.09 
\enddata
\tablenotetext{a}{$\Delta$ is the $[\mathrm{Fe/H}]$ values in \citet{luc11} minus published data sets.}
\tablenotetext{b}{$\sigma$ is the dispersion of the mean value.}
\tablenotetext{c}{Two discrepant Cepheids, CI Per and OT Per, are excluded in the calculation.}
\tablenotetext{d}{A discrepant Cepheid, EV SCT, is excluded in the calculation.}
\end{deluxetable}

We first compared our distances to the Cepheids that have {\it Hipparcos} parallax measurements. Comparison of the parallaxes, measured in milli-arcsecond (mas), for $223$ common Cepheids listed in \citet{van07} and Table \ref{tab1} is shown in left panel of Figure \ref{fig_hipp}. For the $14$ Cepheids listed in Table 2 of \citet{van07}, right panel of Figure \ref{fig_hipp} presents the comparison of {\it Hipparcos} parallaxes and the parallaxes based on distance moduli calculated using equation (1). \citet{van07} has pointed out that Y Sgr shows a discrepancy between the {\it Hipparcos} parallax and {\it HST} parallax given in \citet{ben07}. Another Cepheid that shows a large discrepancy between {\it Hipparcos} and {\it HST} parallax is RT Aur ($-0.23$ mas versus $2.40$ mas). Our parallaxes of $2.05\pm0.08$ mas and $2.13\pm0.08$ mas for Y Sgr and RT Aur, respectively, favor the parallax measured from {\it HST}. Two additional Cepheids, $\beta$ Dor and T Vul, show a difference of $0.58$ mas between {\it Hipparcos} parallaxes and our parallaxes. These four Cepheids were excluded in comparison. On the other hands, excellent agreement has been found between {\it Hipparcos} ($7.72\pm0.12$ mas) and our parallax ($7.71\pm0.29$ mas) for Polaris ($\alpha$ UMi). Parallaxes from {\it Hipparcos} were converted to distance moduli with Lutz-Kelker corrections given in Table 2 of \citet{van07}, and compared to the distance moduli given in Table \ref{tab1}. The weighted mean difference\footnote{Throughout the paper, difference in distance is referred as published distance minus the distance given in Table \ref{tab1}.} of these 10 Cepheids is $-0.014\pm0.056$, with a standard deviation ($\sigma$) of $0.229$.

\begin{figure*}
$\begin{array}{cccc}
\includegraphics[angle=0,scale=0.4]{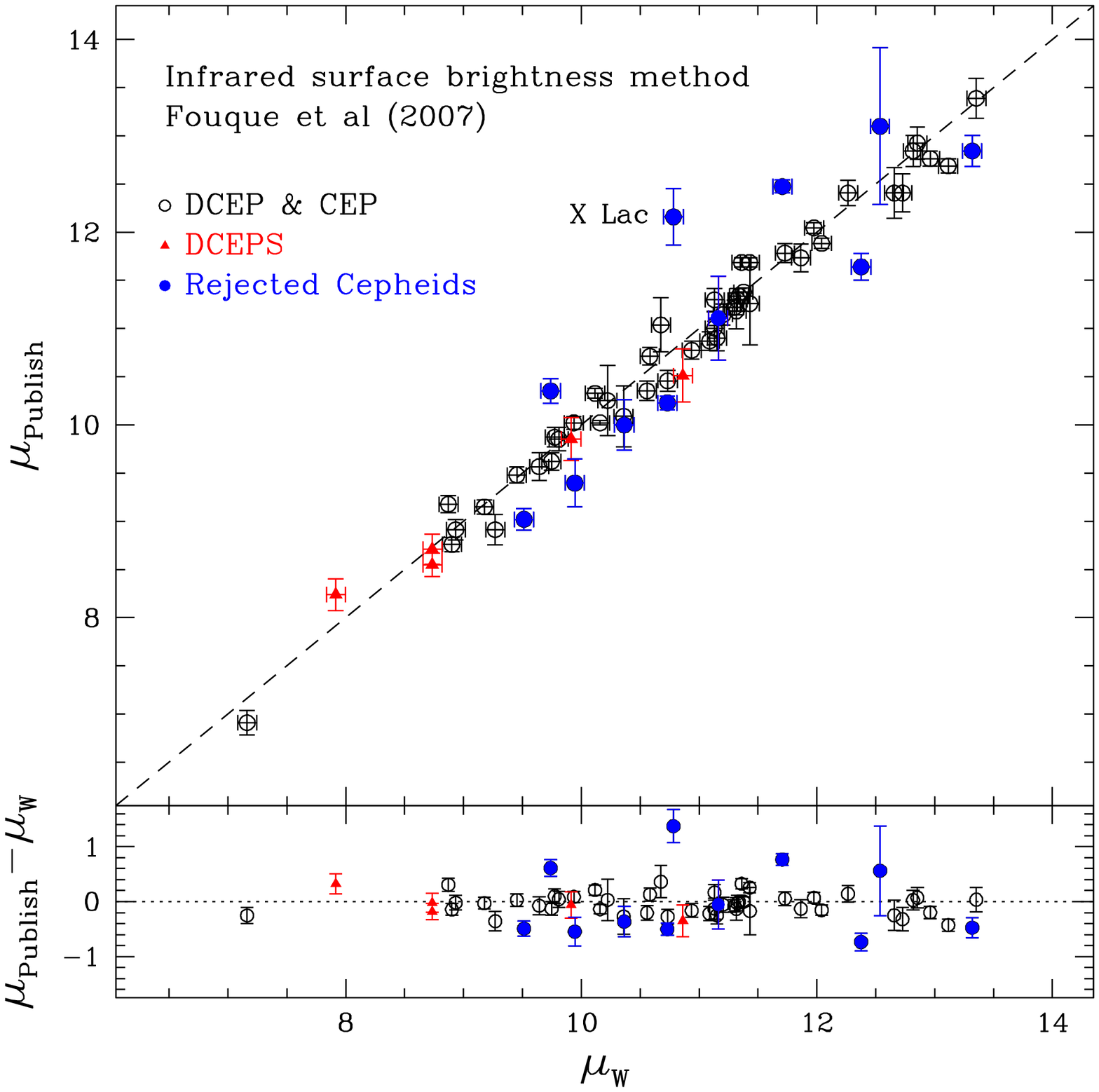} &
\includegraphics[angle=0,scale=0.4]{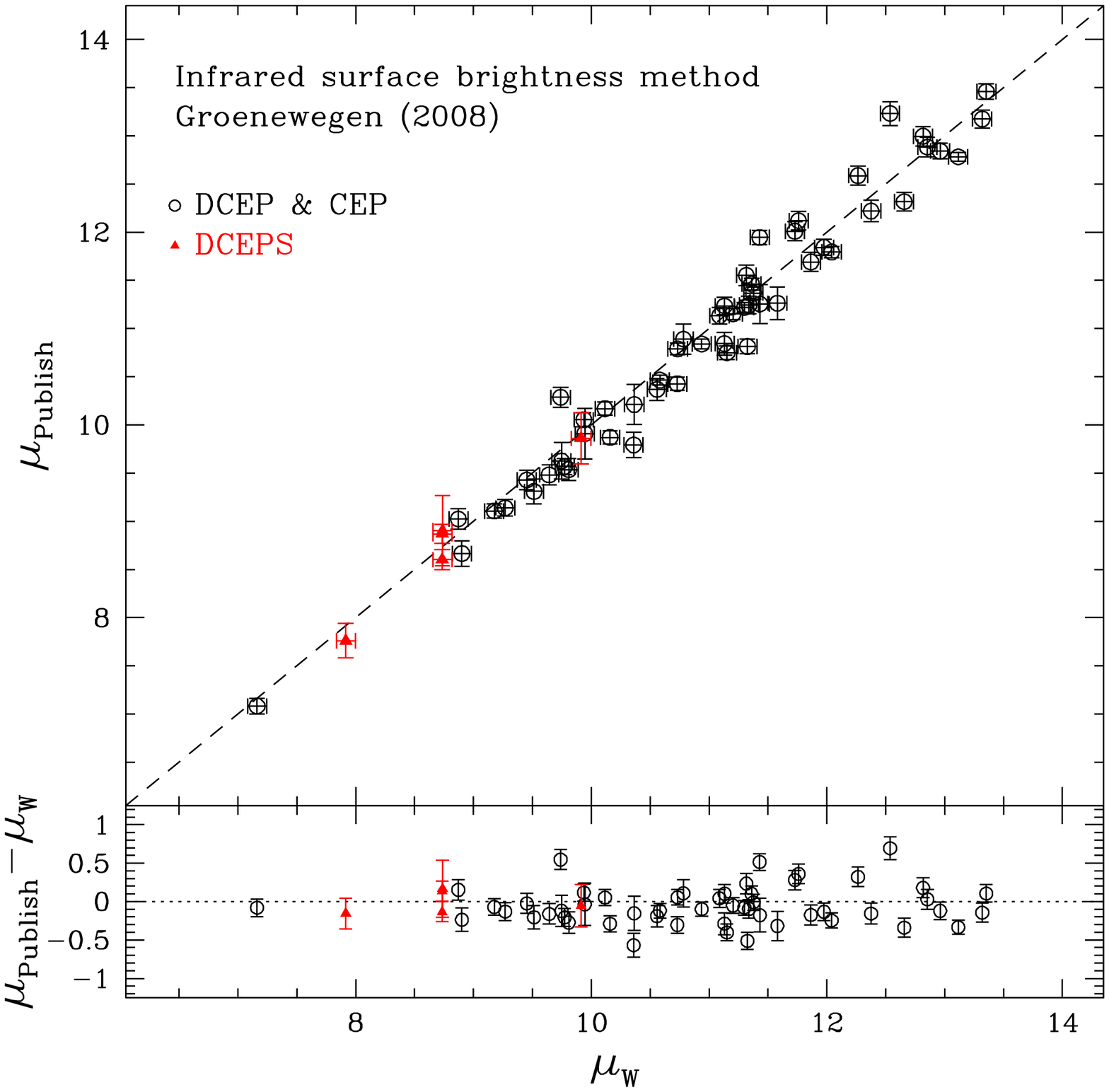} \\
\includegraphics[angle=0,scale=0.4]{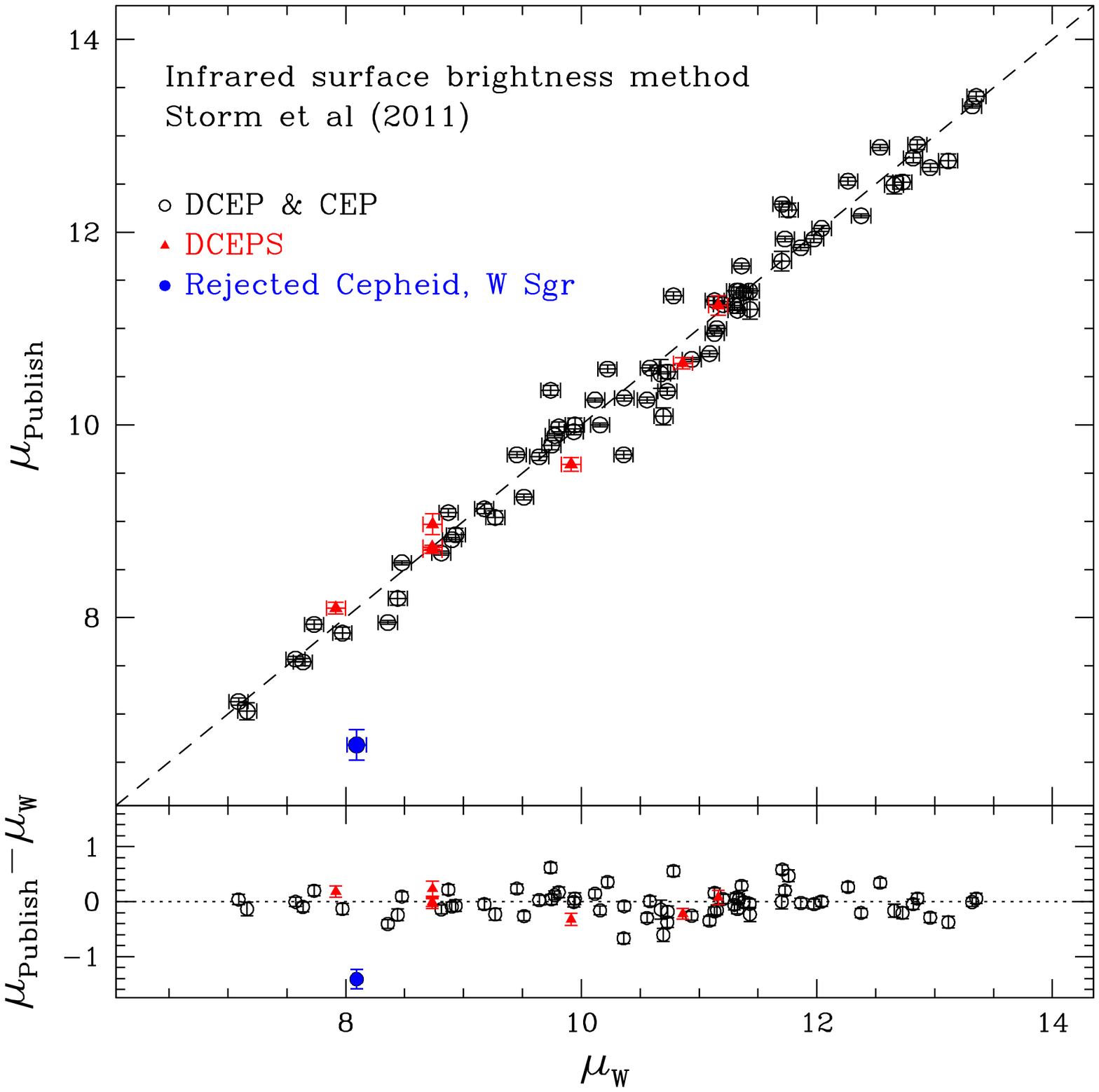} &
\includegraphics[angle=0,scale=0.4]{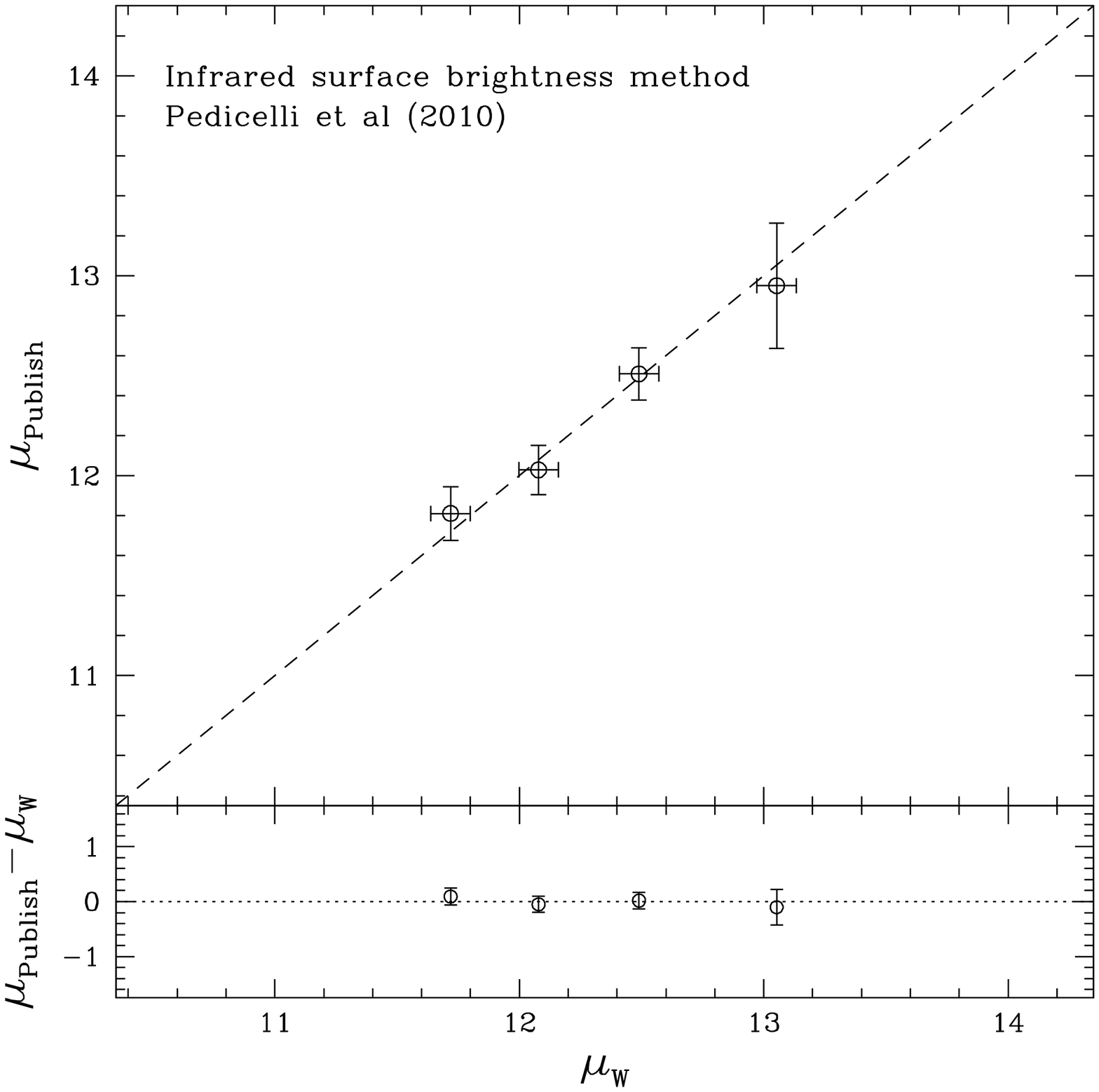} \\
\end{array}$ 
\caption{Comparison of the distance moduli based on IRSB method with distance moduli measured using equation (1) for four different IRSB samples. DCEP and CEP Cepheids are represented by open circles, while DCEPS Cepheids are represented by filled triangles. Excluded Cepheids are labeled with filled circles. The dashed and dotted lines in upper and lower panel are for the case of $y=x$ and $y=0$, respectively, and not the fit to the data. [See on-line edition for color version of this Figure.] \label{fig_bw}}
\end{figure*} 

\begin{figure}
\plotone{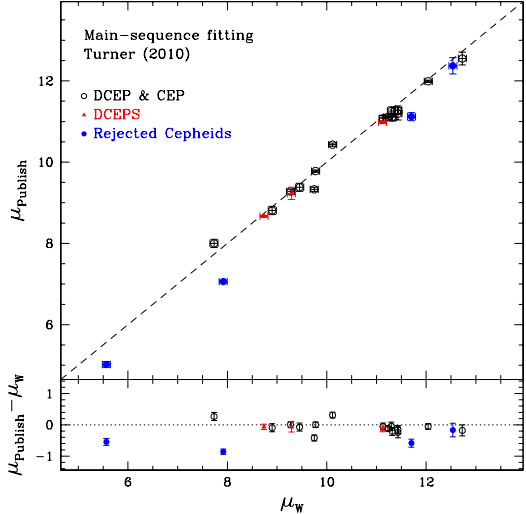}
\caption{Comparison of the distance moduli based on main-sequence fitting method compiled in \citet{tur10} with distance moduli  measured using equation (1). DCEP and CEP Cepheids are represented by open circles, while DCEPS Cepheids are represented by filled triangles. Excluded Cepheids are labeled with filled circles. The dashed and dotted lines in upper and lower panel are for the case of $y=x$ and $y=0$, respectively, and not the fit to the data. [See on-line edition for color version of this Figure.] \label{fig_tur10}}
\end{figure} 

Latest Cepheid distances using BW infrared surface brightness (IRSB) method have been published in \citet{fou07}, \citet{gro08} and \citet{sto11}. For \citet{fou07} sample, 10 Cepheids with low quality in distance measurements or being rejected by \citet{fou07} were excluded\footnote{These Cepheids are: FM Aql, FN Aql, GT Car, BF Oph, X Pup, VZ Pup, GY Sge, YZ Sgr, SS Sct and S Vul.}. X Lac was also excluded as the distance modulus for this Cepheid ($12.159\pm0.293$) is very different from the distance modulus given in Table \ref{tab1} ($10.783$) or from \citet[][$10.891\pm0.157$]{gro08}. The weighted mean difference of the distance moduli for this sample is $-0.025\pm0.019$ ($\sigma=0.190$). For \citet{gro08} sample, the weighted mean difference is $-0.055\pm0.016$ ($\sigma=0.244$). For \citet{sto11} sample, after excluding W Sgr \citep[which is also excluded in][]{sto11} that shows a large difference between the {\it HST} parallax distance and the distance from IRSB, the weighted mean difference is $-0.018\pm0.011$ ($\sigma=0.242$). Finally, IRSB distances to four metal rich Cepheids are available from \citet{ped10}, with a negligible weighted mean difference of $0.009\pm0.085$ ($\sigma=0.088$). Comparisons of the distance moduli for these four samples are shown in Figure \ref{fig_bw}.

\begin{figure}
\epsscale{0.9}
\plotone{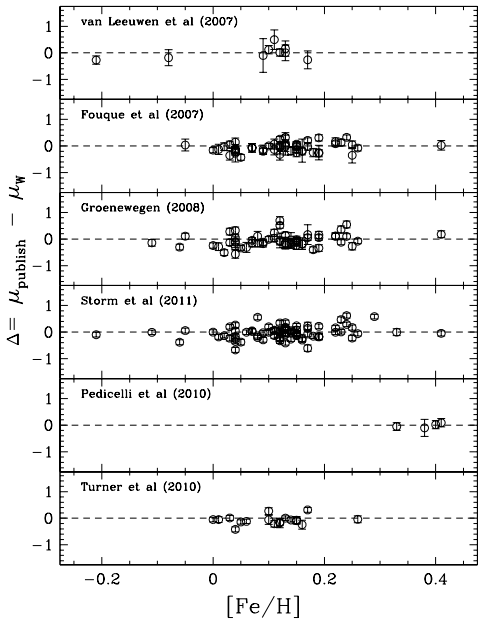}
\caption{Difference in distance moduli for individual Cepheids as a function of $[\mathrm{Fe/H}]$, for the six difference samples considered in this work when comparing the published distance moduli to the distance moduli calculated from equation (1). The dashed horizontal lines are cases for $y=0$ and not the fit to the data. \label{fig_feh}}
\end{figure} 

Recent Cepheid distance measurements based on main-sequence fitting to open clusters and associations that host Cepheids have been compiled in \citet{tur10}. Three Cepheids ($\alpha$ UMi, SU Cas \& S Vul) that show discrepancy between distances from main-sequence fitting and other distance indicators, either from {\it Hipparcos} parallax (for $\alpha$ UMi) or IRSB technique, were excluded. TW Nor was further rejected as the difference in distance modulus from Table \ref{tab1} and \citet{tur10} is more than $0.5$ mag. For the remaining Cepheids, the weighted mean difference in distance moduli is $-0.061\pm0.025$ ($\sigma=0.167$). Figure \ref{fig_tur10} presents the comparison of the distance moduli for this sample of Cepheids.

Figure \ref{fig_feh} shows the difference in distance moduli as a function of $[\mathrm{Fe/H}]$ for individual Cepheids from the samples considered previously. No obvious dependence has been found between the difference in distance moduli, calculated from equation (1) and the published results, and the metallicity for individual Cepheids.

Using a sample of Cepheids that have $[\mathrm{Fe/H}]$ from Table \ref{tab1}, it is straight forward to derive the Galactic metallicity gradient. The Galactocentric distances were calculated using $R^2_G =  R^2_{\odot} + (d\cos b)^2 - 2 d R_{\odot}\cos b \cos l$, where $d=10^{0.2\mu_W-1}$ is the distance to Cepheids in $\mathrm{kpc}$ (with $\mu_W$ taken from Table \ref{tab1}), $l$ and $b$ are Galactic coordinates in radians, and $R_{\odot}=7.9\mathrm{kpc}$ is the Galactocentric of the Sun \citep{mcn00}. The resulting metallicity gradient is: $[\mathrm{Fe/H}]=0.580(\pm0.024)-0.058(\pm0.003) R_G$, with a dispersion of $0.101$, which is consistent with the relation found by \citet{luc11}. 

\section{The Galactic Period-Luminosity Relations and Wesenheit Functions}

\begin{figure*}
\plottwo{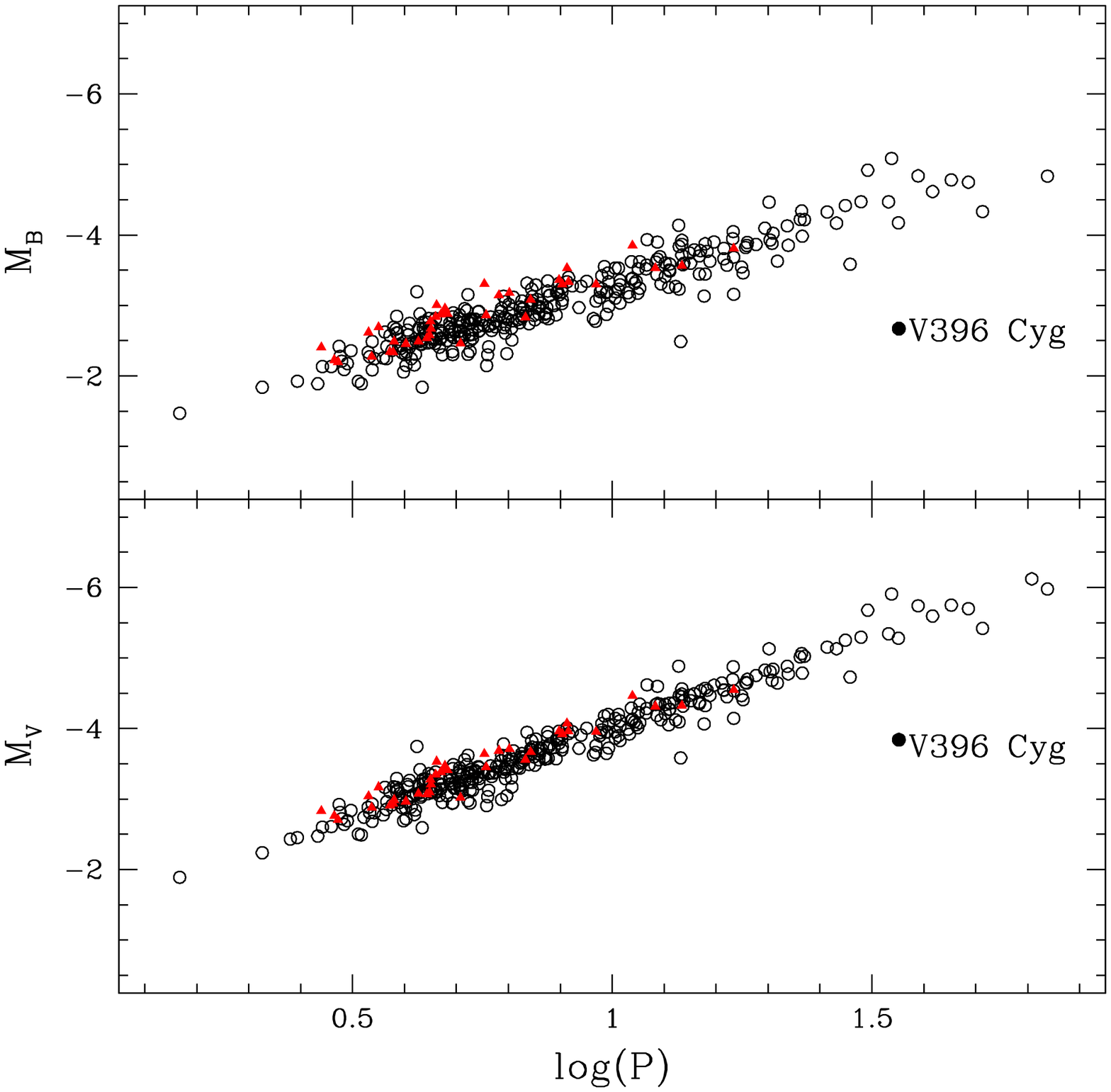}{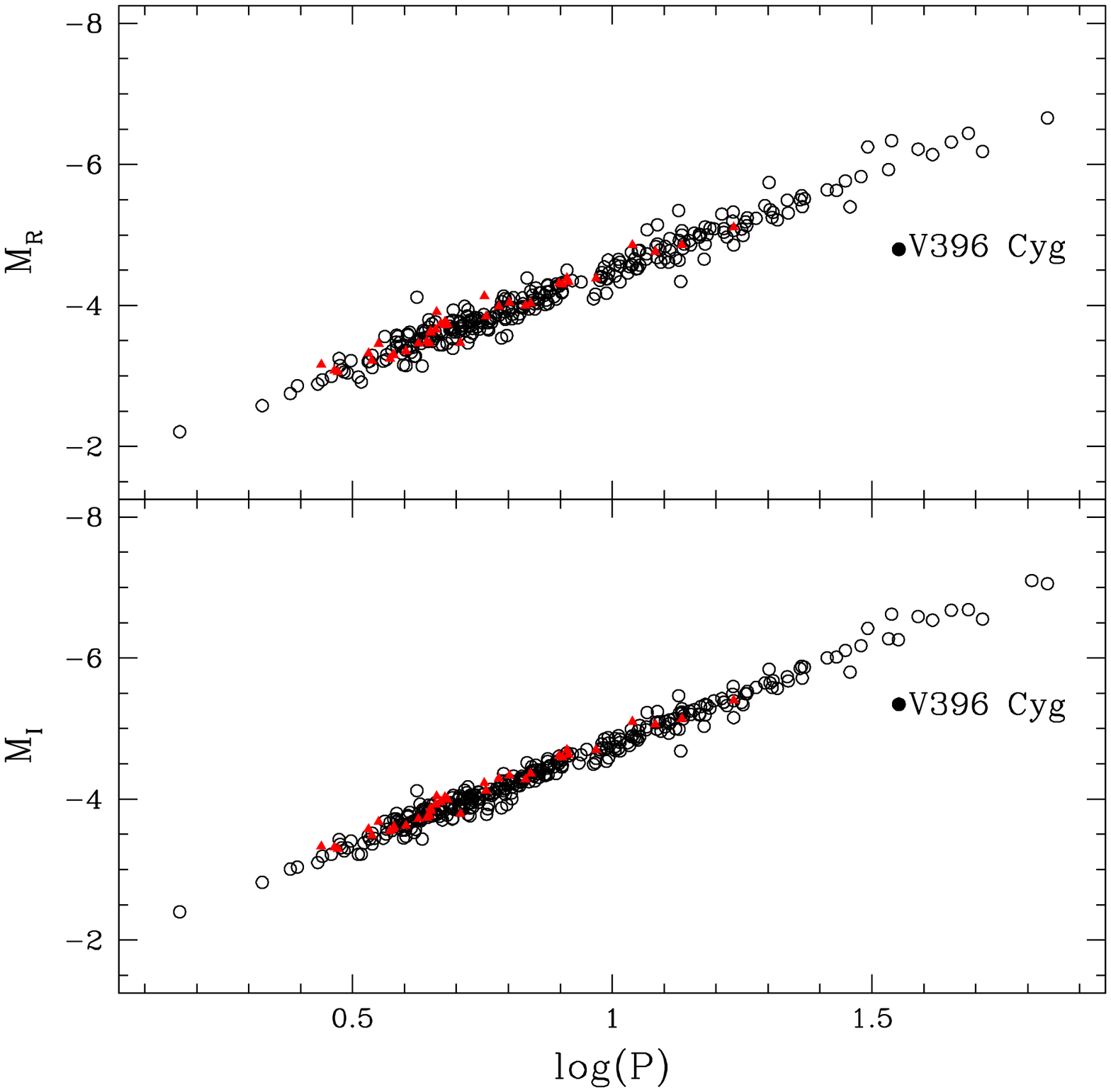}
\caption{Extinction corrected $BVRI$ band P-L relations, based on the mean intensities and distance moduli listed in Table \ref{tab1}. Open circles are for Cepheids classified as DCEP and CEP, and filled (red) triangles are for DCEPS Cepheids. The outlier is marked as filled circles. [See on-line edition for color version of this Figure.] \label{fig_pl}}
\end{figure*} 

\begin{figure}
\plotone{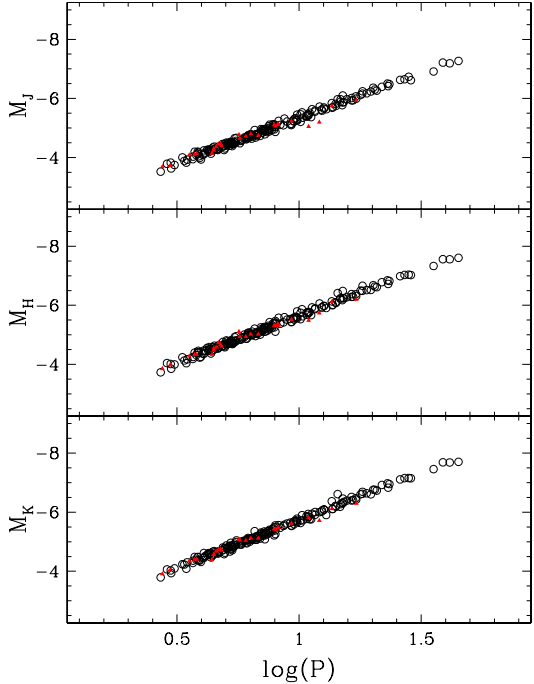}
\caption{Same as Figure \ref{fig_pl}, but for $JHK$ bands in 2MASS system. [See on-line edition for color version of this Figure.] \label{fig_jhk}}
\end{figure} 

\begin{deluxetable}{lrccc}
\tabletypesize{\scriptsize}
\tablecaption{The Galactic P-L Relations\tablenotemark{a}. \label{tab_pl}}
\tablewidth{0pt}
\tablehead{
\colhead{Band} &
\colhead{$N$} & 
\colhead{$a_{\lambda}$} &
\colhead{$b_{\lambda}$} &
\colhead{$\sigma$} 
}
\startdata
  & \multicolumn{4}{c}{All Cepheids} \\
$B$                 & 357 & $-2.103\pm0.044$ & $-1.201\pm0.039$ & 0.221  \\
$V$                 & 364 & $-2.513\pm0.033$ & $-1.508\pm0.030$ & 0.173  \\
$R$                 & 319 & $-2.722\pm0.029$ & $-1.781\pm0.026$ & 0.143  \\
$I$                 & 364 & $-2.826\pm0.020$ & $-1.951\pm0.018$ & 0.105  \\ 
$J$                 & 229 & $-3.030\pm0.022$ & $-2.306\pm0.020$ & 0.082  \\
$H$                 & 229 & $-3.166\pm0.022$ & $-2.478\pm0.020$ & 0.080  \\
$K$                 & 229 & $-3.217\pm0.021$ & $-2.513\pm0.019$ & 0.077  \\
$3.6\mu \mathrm{m}$ &  29 & $-3.242\pm0.067$ & $-2.491\pm0.070$ & 0.115  \\
$4.5\mu \mathrm{m}$ &  29 & $-3.180\pm0.070$ & $-2.523\pm0.073$ & 0.120  \\
$5.8\mu \mathrm{m}$ &  29 & $-3.216\pm0.072$ & $-2.480\pm0.074$ & 0.123  \\ 
$8.0\mu \mathrm{m}$ &  29 & $-3.280\pm0.068$ & $-2.447\pm0.071$ & 0.117  \\
$24\mu \mathrm{m}$  &  29 & $-3.341\pm0.062$ & $-2.420\pm0.064$ & 0.107  \\
  & \multicolumn{4}{c}{Exclude DCEPS} \\
$B$                 & 320 & $-2.132\pm0.045$ & $-1.159\pm0.041$ & 0.219  \\
$V$                 & 327 & $-2.534\pm0.035$ & $-1.478\pm0.032$ & 0.173  \\
$R$                 & 282 & $-2.734\pm0.031$ & $-1.764\pm0.029$ & 0.145  \\
$I$                 & 327 & $-2.839\pm0.021$ & $-1.932\pm0.019$ & 0.105  \\
$J$                 & 203 & $-3.058\pm0.021$ & $-2.282\pm0.019$ & 0.073  \\
$H$                 & 203 & $-3.181\pm0.022$ & $-2.467\pm0.020$ & 0.077  \\
$K$                 & 203 & $-3.231\pm0.021$ & $-2.501\pm0.020$ & 0.075  \\
$3.6\mu \mathrm{m}$ &  24 & $-3.289\pm0.065$ & $-2.454\pm0.069$ & 0.103  \\
$4.5\mu \mathrm{m}$ &  24 & $-3.233\pm0.072$ & $-2.472\pm0.077$ & 0.114  \\
$5.8\mu \mathrm{m}$ &  24 & $-3.277\pm0.065$ & $-2.420\pm0.070$ & 0.104  \\
$8.0\mu \mathrm{m}$ &  24 & $-3.337\pm0.061$ & $-2.395\pm0.065$ & 0.096  \\
$24\mu \mathrm{m}$  &  24 & $-3.366\pm0.054$ & $-2.413\pm0.058$ & 0.086  \\
  & \multicolumn{4}{c}{DCEPS Only} \\
$B$    & 37  & $-2.168\pm0.158$ & $-1.302\pm0.120$ & 0.182  \\
$V$    & 37  & $-2.511\pm0.120$ & $-1.610\pm0.091$ & 0.139  \\
$R$    & 37  & $-2.728\pm0.101$ & $-1.822\pm0.076$ & 0.116  \\
$I$    & 37  & $-2.824\pm0.073$ & $-2.013\pm0.056$ & 0.085  \\
$J$    & 26  & $-2.704\pm0.118$ & $-2.562\pm0.095$ & 0.120  \\
$H$    & 26  & $-2.945\pm0.095$ & $-2.634\pm0.077$ & 0.097  \\
$K$    & 26  & $-3.015\pm0.083$ & $-2.661\pm0.067$ & 0.085  
\enddata
\tablenotetext{a}{The P-L relation takes the form of $M_{\lambda}=a_{\lambda}\log(P)+b_{\lambda}$, $\sigma$ is the dispersion of the fitted P-L relations.}
\end{deluxetable}

\begin{figure*}
\plottwo{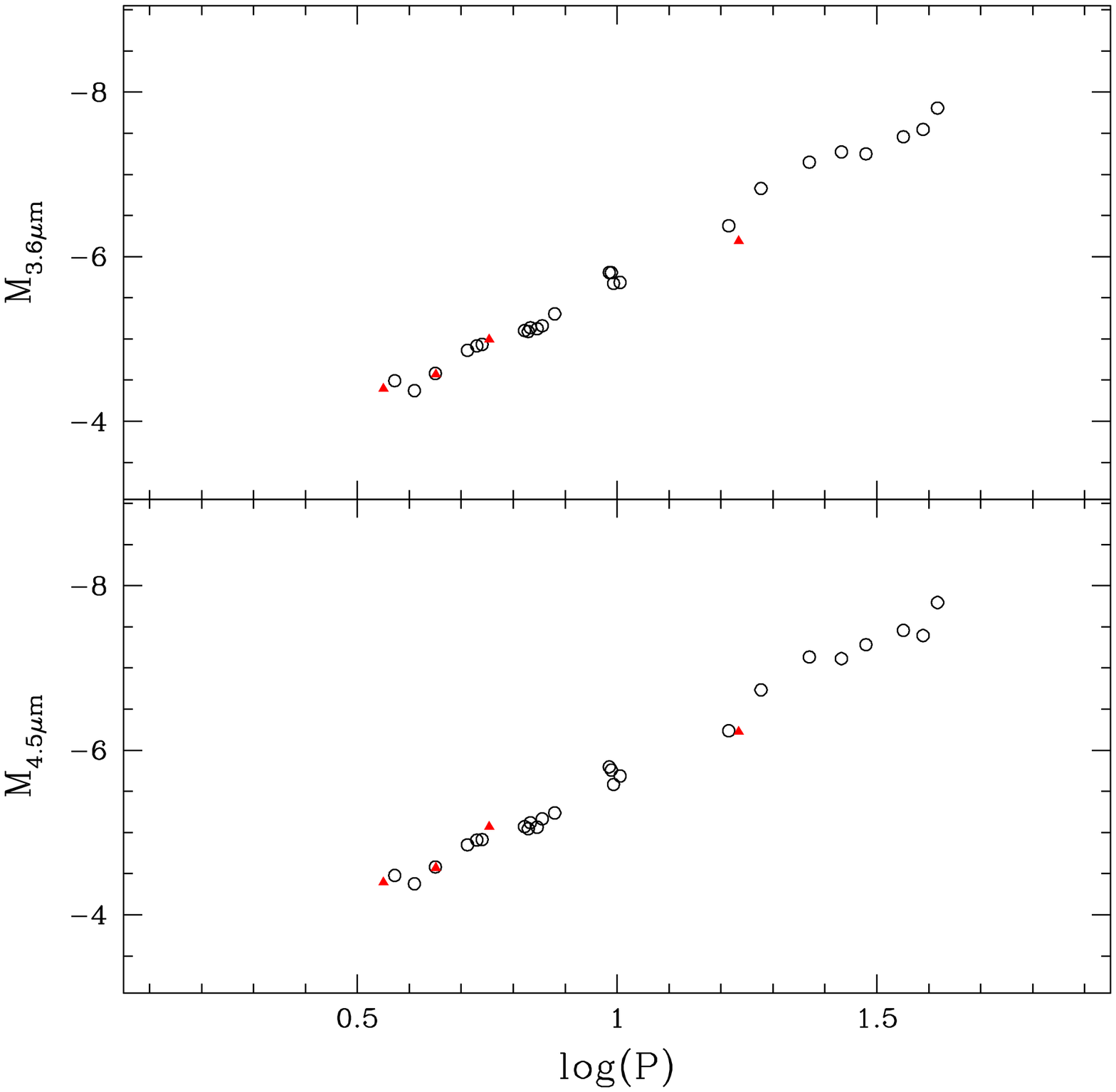}{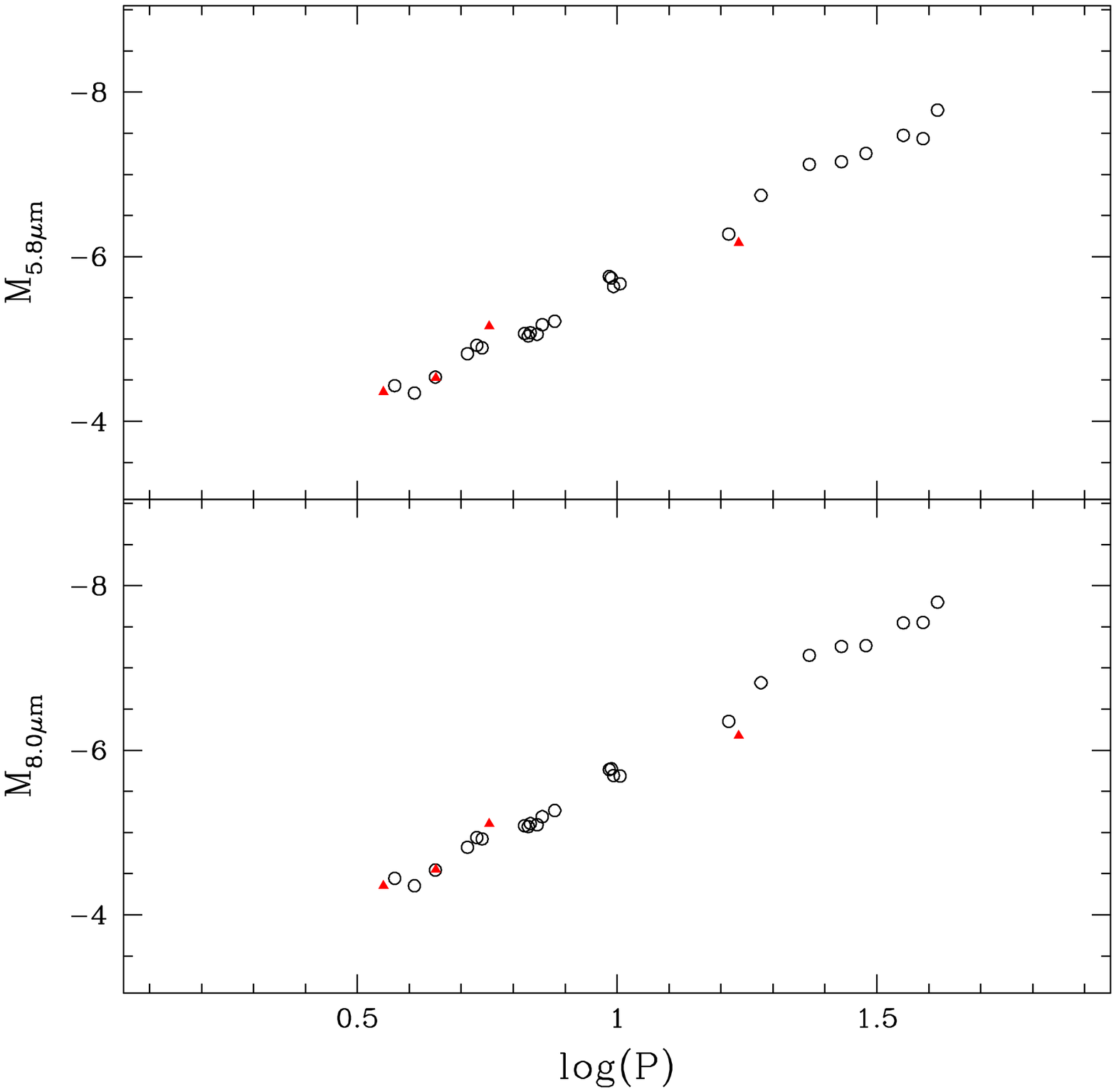}
\caption{Same as Figure \ref{fig_pl}, but for {\it Spitzer} IRAC bands. [See on-line edition for color version of this Figure.] \label{fig_irac}}
\end{figure*}

\begin{figure}
\plotone{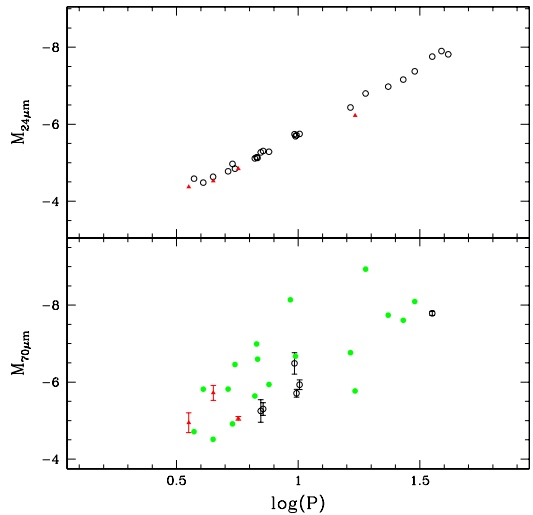}
\caption{Same as Figure \ref{fig_pl}, but for {\it Spitzer} MIPS bands. Cepheids that only have the upper flux limits (lower limits in magnitudes) in MIPS $70\mu \mathrm{m}$ are represented as filled (green) circles. [See on-line edition for color version of this Figure.] \label{fig_mips}}
\end{figure}

It is straight forward to derive the Galactic P-L relations using the data and distance moduli given in Table \ref{tab1}. For the 28 Cepheids that do not have $E(B-V)$ values are not used in deriving the P-L relations. Total-to-selective absorption ratios were adopted from \citet{fou07} as follow: $R_{\{B,\ V,\ R,\ I,\ J,\ H,\ K\}}=\{4.23,\ 3.23,\ 2.73,\ 1.96,\ 0.94,\ 0.58,\ 0.38\}$. Finally, {\it Spitzer} IRAC and MIPS band photometry were available from \citet{mar10} for 29 Cepheids. Extinction is ignored for the mid-infrared band photometry, as it is negligible at these wavelengths \citep{fre08,nge08,nge09,fre10,fre11}. The $70\mu \mathrm{m}$ band photometric data are not considered in this work, as most of them only have upper limits in flux. The multi-band P-L relations are presented in Figure \ref{fig_pl} to \ref{fig_mips}.

\begin{figure*}
\plottwo{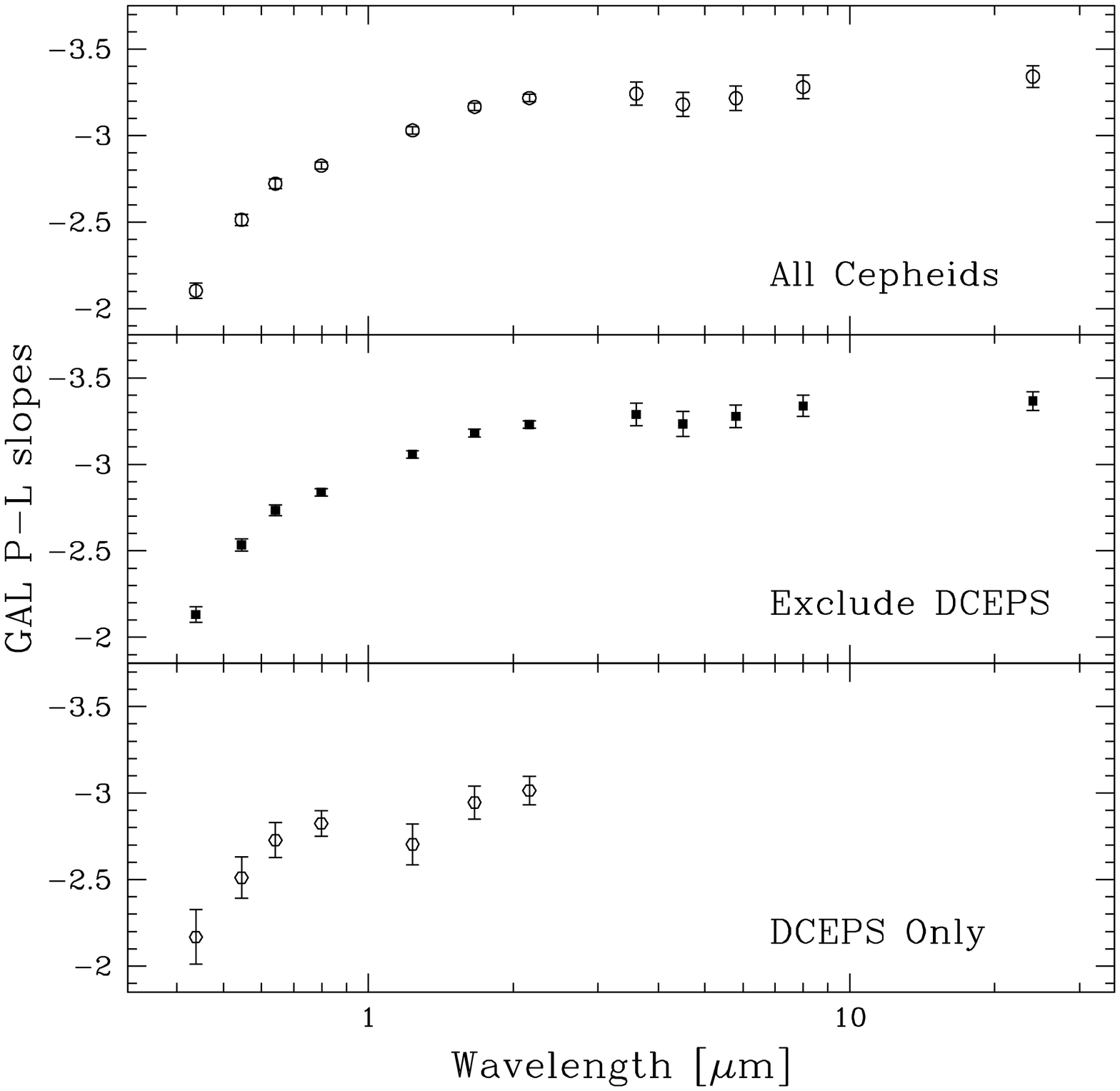}{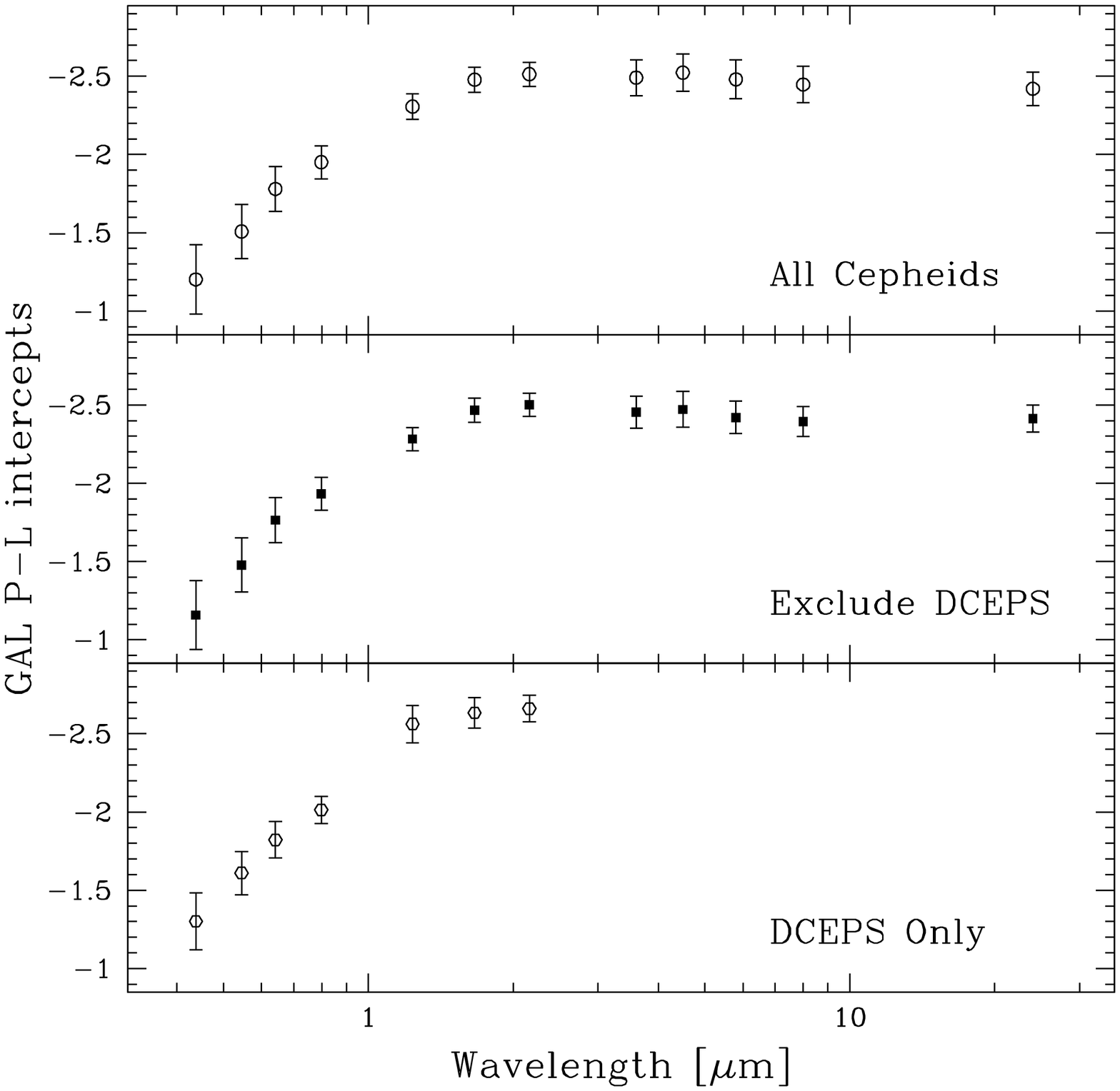}
\caption{P-L slopes (left panels) and intercepts (right panels) as a function of wavelength based on the results presented in Table \ref{tab_pl}. \label{fig_wavelength}}
\end{figure*}

A clear outlier, V396 Cyg, is shown up in Figure \ref{fig_pl}. Based on its location in P-L plane, it is possible that this Cepheid is a type II Cepheid, and not a classical population I Cepheid. Wesenheit function can be used to disentangle type II Cepheids from the classical Cepheids \citep[for example, see][their Figure 1]{sos08b}. For Wesenheit functions in the form of $W_{BV}=V-3.23(B-V)$ and $W_{RI}=I-2.55(R-I)$, the Wesenheit magnitudes of this outlier is $\sim0.69\mathrm{mag}(=4.1\sigma$, where $\sigma$ is the dispersion of the fitted Wesenheit function) and $\sim0.36\mathrm{mag}(=2.7\sigma)$, respectively, fainter from the ridge line of the fitted Wesenheit function given in Table \ref{tabw}. Therefore, it is inconclusive whether this outlier should be type II or classical Cepheid. Another possible explanation is that the published extinction value, $E(B-V)=1.092$, for this outlier is underestimated. A higher value of $E(B-V)\sim1.5$ can bring a better agreement for its absolute magnitudes to other Cepheids with similar periods. Detailed investigation of this Cepheid is beyond the scope of this paper, but nevertheless it is clear that this Cepheid should be excluded when fitting the P-L relations. For remaining Cepheids, the fitted multi-band P-L relations are summarized in Table \ref{tab_pl}. Slopes and intercepts of these P-L relations as a function of wavelengths were presented in Figure \ref{fig_wavelength}. Both the P-L slopes and intercepts monotonically decrease from $B$ to $K$ band (except for $J$ band P-L slope with DCEPS Cepheids), and approach an asymptotic value in the mid-infrared. The ``dip'' around $4.5\mu \mathrm{m}$ and $5.8\mu \mathrm{m}$ band P-L slopes is interpreted as due to the CO absorption that affecting this band \citep{mar10}.

\begin{deluxetable}{lcccr}
\tabletypesize{\scriptsize}
\tablecaption{Selected Galactic Wesenheit Functions\tablenotemark{a}. \label{tabw}}
\tablewidth{0pt}
\tablehead{
\colhead{$W=$} &
\colhead{$a$} & 
\colhead{$b$} &
\colhead{$\sigma$} &
\colhead{$N$} 
}
\startdata
 \multicolumn{5}{c}{All Cepheids} \\
$V-3.23(B-V)$  & $-3.811\pm0.033$ & $-2.514\pm0.030$ & 0.179 & 384 \\
$I-2.55(R-I)$  & $-3.091\pm0.025$ & $-2.394\pm0.023$ & 0.130 & 333 \\
$J-1.68(J-K)$  & $-3.343\pm0.026$ & $-2.654\pm0.024$ & 0.097 & 229 \\
$J-0.74(V-I)$  & $-3.276\pm0.019$ & $-2.615\pm0.017$ & 0.071 & 229 \\
$H-0.46(V-I)$  & $-3.320\pm0.021$ & $-2.672\pm0.019$ & 0.077 & 229 \\
$K-0.30(V-I)$  & $-3.317\pm0.020$ & $-2.639\pm0.018$ & 0.074 & 229 \\
$H-0.41(V-I)$  & $-3.293\pm0.020$ & $-2.633\pm0.019$ & 0.076 & 229 \\
 \multicolumn{5}{c}{Exclude DCEPS} \\
$V-3.23(B-V)$  & $-3.811\pm0.034$ & $-2.520\pm0.032$ & 0.178 & 347 \\
$I-2.55(R-I)$  & $-3.113\pm0.026$ & $-2.362\pm0.024$ & 0.128 & 296 \\
$J-1.68(J-K)$  & $-3.350\pm0.027$ & $-2.650\pm0.025$ & 0.097 & 203 \\
$J-0.74(V-I)$  & $-3.297\pm0.016$ & $-2.600\pm0.015$ & 0.057 & 203 \\
$H-0.46(V-I)$  & $-3.330\pm0.020$ & $-2.667\pm0.019$ & 0.071 & 203 \\
$K-0.30(V-I)$  & $-3.328\pm0.020$ & $-2.631\pm0.019$ & 0.071 & 203 \\
$H-0.41(V-I)$  & $-3.307\pm0.020$ & $-2.624\pm0.018$ & 0.071 & 203 \\
 \multicolumn{5}{c}{DCEPS Only} \\
$V-3.23(B-V)$  & $-3.620\pm0.154$ & $-2.605\pm0.117$ & 0.177 & 37  \\
$I-2.55(R-I)$  & $-3.072\pm0.094$ & $-2.500\pm0.071$ & 0.108 & 37  \\
$J-1.68(J-K)$  & $-3.227\pm0.100$ & $-2.729\pm0.081$ & 0.102 & 26  \\
$J-0.74(V-I)$  & $-2.955\pm0.120$ & $-2.839\pm0.097$ & 0.122 & 26  \\
$H-0.46(V-I)$  & $-3.103\pm0.094$ & $-2.806\pm0.094$ & 0.096 & 26  \\
$K-0.30(V-I)$  & $-3.117\pm0.078$ & $-2.773\pm0.063$ & 0.080 & 26  \\
$H-0.41(V-I)$  & $-3.059\pm0.096$ & $-2.790\pm0.077$ & 0.098 & 26  
\enddata
\tablenotetext{a}{The Wesenheit function takes the form of $W=a\log(P)+b$, $\sigma$ is the dispersion of the fitted Wesenheit functions.}
\end{deluxetable}

Data in Table \ref{tab1} can also be used to derive the Galactic Wesenheit functions in other bandpass and color combinations (except for the $VI$ bands). A number of selected Wesenheit functions involving $BRJHK$ bands are presented in Table \ref{tabw}.\footnote{Other forms of Wesenheit functions is straight forward to derive from Table \ref{tab1}, hence they are not included here.} Wesenheit function in the from of $W=H-0.41(V-I)$, adopted by the SH0ES team \citep{rie11}, is also included in Table \ref{tabw}. For Wesenheit function involved $B$ band, the dispersion of the relation is largest with steepest slope, which is consistent with the results found in \cite{bon10}. The dispersions of the $JHK$ band based Wesenheit functions, on the other hand, are in the order of $\sim0.1$ and smaller, suggesting they could also be used in future distance scale work. It is worth to point out that Wesenheit function in the form of $W=J-0.74(V-I)$, based on DCEP and CEP Cepheids, has a dispersion of $0.057$, about $17$\% smaller than the $VI$ band based Wesenheit function adopted in this work.

\section{Discussion and Conclusion}

In conclusion, mean differences between the distance moduli given in literature and those calculated from equation (1) range from negligible to about $6$ per-cent, depending on the samples and the methods used to derive independent distances to the Galactic Cepheids. However, various assumptions have been made when deriving these independent distances. For example, certain expression of the $p$-factors (to covert radial velocities to pulsational velocities) need to be adopted when applying the BW type analysis. In contrast, equation (1) is assumed to be applicable to Galactic Cepheids, though the relation is derived from LMC Cepheids. Good agreement between the 10 Cepheids with {\it Hipparcos} parallaxes and the distance calculated from equation (1) suggested that the Wesenheit function used in this work is not sensitive, or mildly sensitive, to metallicity. This is also supported by recent empirical work presented in \citet[][and reference therein]{bon10}, showing the empirical slopes of the Wesenheit functions for both metal poor and metal rich galaxies are consistent with the LMC Wesenheit slopes. \citet{maj11} have also demonstrated that the distance moduli to Magellanic Clouds would be significantly disagreement with the canonical values if adopting a strong metallicity correction to $VI$ band based Wesenheit function. In terms of theoretical works, \citet{fio07} and \citet{bon08} also found a weak or mild dependence of $VI$ band based Wesenheit function on metallicity. Other advantages of using the $VI$ band based Wesenheit function, in addition to being reddening-free by definition, include the smaller dispersion of the relation, being linear \citep{mar05,nge05,fio07,mad09,nge09,bon10}, and only need the $VI$ band mean magnitudes and periods of the Cepheids. Certainly, the verification of the distance moduli given in Table \ref{tab1} and the use of Wesenheit function in deriving distances relies on the parallaxes measured from {\it Gaia}.

\acknowledgments

The author thanks Dr. Shashi Kanbur and referee for useful comments to improve this manuscript. CCN thanks the funding from National Science Council (Taiwan) under the contract NSC 98-2112-M-008-013-MY3. 


\end{document}